\input{aipcheck}
\newcommand{\cz}{\ensuremath{C_Z}}
\newcommand{\pv}{\ensuremath{P_V}}
\newcommand{\nv}{\ensuremath{N_V}}

\newcommand{\bz}{\ensuremath{\langle B_z\rangle}}
\newcommand{\sbz}{\ensuremath{\sigma_{\langle B_z\rangle}}}
\newcommand{\nz}{\ensuremath{\langle N_z\rangle}}

\newcommand{\fo}{\ensuremath{f^\parallel}}
\newcommand{\fe}{\ensuremath{f^\perp}}

\documentclass[final]{aipproc}
\layoutstyle{6x9}

\begin{document}

\title{Stellar magnetism through the polarized eyes\\ of the FORS1 instrument}


\author{S. Bagnulo}{
  address={Armagh Observatory, College Hill, Armagh BT61 9DG, UK}
}

\author{J.D. Landstreet}{
  address={Physics \& Astronomy Department,
           The University of Western Ontario,
           London, Ontario, Canada N6A 3K7.}
}

\author{L. Fossati}{
  address={Department of Physics and Astronomy, Open University,
           Walton Hall, Milton Keynes MK7 6AA} 
}

\author{O. Kochukhov}{
  address={Department of Physics and Astronomy,
           Uppsala University,
           751 20 Uppsala, Sweden}
}

\begin{abstract}
During the last decade, the FORS1 instrument of the ESO Very Large
Telescope has been used to obtain low resolution circular polarized
spectra for about 500 stars, with the aim of measuring their mean
longitudinal magnetic fields. Magnetic field estimates were obtained
by different authors, using different software tools.  Several
interesting detections were obtained at a $3\,\sigma$ level; some of
them were eventually confirmed by follow-up investigations, some of
them were not. This raises issues about the reliability of the stated
uncertainties of some of the published field values. To investigate
these problems, we have developed a semi-automatic procedure for
magnetic field determination, which includes self-consistent checks
for field detection reliability. We have applied our procedure to the
full content of single star ("fast mode") circular spectropolarimetric
measurements of the FORS1 archive, and explored the details and
interagreement of various methods for data reduction. We have finally
produced a catalogue of FORS1 longitudinal field measurements which
includes about 1000 entries. Here we critically review the previously
published FORS1 measurements, and, based on our results, we suggest that
the incidence of the magnetic field in various classes of stars should
be revised.
\end{abstract}

\maketitle

\section{1. Introduction}\label{Sect_Introduction}
The FORS1 instrument (Appenzeller et al.\ 1998) of the ESO VLT has been
extensively used for magnetic field measurements in various classes of
stars. The large majority of FORS1 measurements have been published in the
literature, and we thought that a general catalogue would serve to
obtain an overview (even though biased on target selection) of the
incidence of the magnetic fields in various kinds of stars. However, a
catalogue compiled using published material would suffer from the lack
of homogeneity in the way data have been treated. Furthermore, over
time, new ideas for data reduction and quality checks have improved
the reliability of FORS1 magnetic measurements, which calls for a
revision of earlier data. We also noted that the literature of FORS
magnetic field measurements includes a certain amount of controversial
detections, e.g.: (i) FORS detections not confirmed by
observations obtained with other instruments; (ii) FORS detections not
confirmed by further FORS measurements; (iii) FORS detections not
confirmed by the re-analysis of the same data by other groups; (iv)
FORS detections not confirmed by the re-analysis of the same data
performed by the same group who originally claimed detection.

The release of the FORS pipeline for spectropolarimetric data (Izzo et
al.  2010, 2011) gave us the opportunity to develop a reduction method
potentially more accurate than what available in the past.
Compared to the reduction of raw data coming from individual observing
runs, the mass-production of reduced spectra offers a few advantages
in terms of quality of the final products, as the results can be
checked on a large statistical basis.  Therefore we have decided to tackle
the task of re-reducing the entire FORS archive of spectropolarimetric
data obtained in single target mode (``fast mode''), and compile a new
catalogue. In this paper we report on some of our most important
findings.

In Sect.~2 and 3 we give details about data reduction and field
measurements techniques, and in Sects.~4 and 5 we present our results
and conclusions.

\section{2. Magnetic field measurements}\label{Sect_Field_Meas}
The mean longitudinal magnetic field is obtained from the Stokes
parameters $I$ and $V$ using the relationship
%%%%%%%%%%%%%%%%
\begin{equation}
\frac{V}{I} = - g_\mathrm{eff} \ \cz \ \lambda^{2} \
                \frac{1}{I} \
                \frac{\mathrm{d}I}{\mathrm{d}\lambda} \
                \bz\;,
\label{Eq_Bz}
\end{equation}
%%%%%%%%%%%%%%%%
where $g_\mathrm{eff}$ is the effective land\'{e} factor, and
%%%%%%%%%%%%%%%%
\begin{equation}
\cz = \frac{e}{4 \pi m_\mathrm{e} c^2}
\ \ \ \ \ (\simeq 4.67 \times 10^{-13} \,{\rm A}^{-1}\ \mathrm{G}^{-1})
\end{equation}
%%%%%%%%%%%%%%%%
where $e$ is the electron charge, $m_\mathrm{e}$ the electron mass,
$c$ the speed of light. It should be reminded that Eq.~(\ref{Eq_Bz})
is subject to several important limitations: (i) it is valid only in
the limit of a magnetic field weak enough that Zeeman splitting is
small compared to the {\em local} spectral line width (i.e., typically
for fields of the order of a kG or less for optical metal spectral
lines, or of order 10\,kG for Balmer lines); (ii) it applies to
isolated, unblended lines; (iii) the value of $g_{\rm eff}$ varies by
a significant amount from line to line; the use of an average value
for all lines means that the actual \bz\ value varies from the
computed value in individual lines by up to $\sim 25$\,\%.

Bagnulo et al.\ (2002) proposed that Eq.~(\ref{Eq_Bz}) could be applied
to spectropolarimetric data obtained with FORS1, using a least-squares
technique that was subsequently extensively adopted in many surveys.
Although field measurements obtained with Eq.~(\ref{Eq_Bz}) cannot be
used for detailed modelling, this technique is certainly valuable
to decide whether a star is magnetic or not, and to estimate the
field strength with a reasonable approximation.

\section{3. FORS1 archive data reduction}\label{Sect_Data_Reduction}
All archive data obtained in fast mode have been pre-processed using
the ESO FORS pipeline (Izzo et al.\ 2010, 2011). Although this
software tool is capable of delivering the final Stokes profiles, we
preferred to develop a few algorithms that further improve the quality
of the results. Therefore, instead of adopting the final products of
the pipeline (i.e., the Stokes profiles), we have used the individual
fluxes corresponding to the parallel and perpendicular beams split by
the Wollaston prism (obtained at the various positions of the retarder
waveplate), and combined them as explained below.

We adopt the same formalism used in Bagnulo et al.\ (2009), i.e., \fo\ and
\fe\ are the fluxes in the parallel and in the perpendicular beam of
the polarization analyser, respectively, $\pv\ = V/I$ is the circular
polarization normalised to the intensity, and \nv\ is the null
profile, a quantity that was introduced by Donati et al.\ (1997) , and
that is representative of the noise of \pv\ (see Bagnulo et
al. 2009). As a quality check, a magnetic field can be deduced from
the null profiles using the same formulas adopted to measure \bz. The
diagnostic content of this \textit{null field} value, \nz, will be
discussed in Sect.~3.2.

The extracted fluxes \fo\ and \fe\ are usually combined to obtain the
\pv\ and \nv\ profiles using the formulas of the difference methods
given in Eqs.~(A2) and (A7) of Bagnulo et al.\ (2009), which for
convenience we reproduce below:
%%%%%%%%%%%%%%%%%%%%%%%%%%%%%%%%%%%%%%%%%%%%%%%%%%%%%%%%%%%%%%%
\begin{equation}
\begin{array}{rcl}
\pv &=& {1 \over 2 N} \sum\limits_{j=1}^N \left[ 
\left(\frac{\fo - \fe}{\fo + \fe}\right)_{\alpha_j} - 
\left(\frac{\fo - \fe}{\fo + \fe}\right)_{\alpha_j + 90^\circ}\right] \\[2mm]
\nv &=& {1 \over 2 N} \sum\limits_{j=1}^N (-1)^{(j-1)}\left[ 
\left(\frac{\fo - \fe}{\fo + \fe}\right)_{\alpha_j} - 
\left(\frac{\fo - \fe}{\fo + \fe}\right)_{\alpha_j + 90^\circ}\right]\; ,\\
\end{array}
\label{Eq_V_and_N}
\end{equation}
%%%%%%%%%%%%%%%%%%%%%%%%%%%%%%%%%%%%%%%%%%%%%%%%%%%%%%%%%%%%%%%
where $\alpha_j$ belongs to the set 
$\{-45^\circ$, $135^\circ \}$. We note that instead of setting
the $\lambda/4$ retarder waveplate to all four possible angles, most
of the observers preferred to set it only to the angles $-45^\circ$ and
$+45^\circ$.

In practice, in many cases we found the \pv\ profile slightly offset
from zero, even when no circular polarization of the continuum was
expected.  A possible explanation is cross-talk from linear to
circular polarization, as discussed by Bagnulo et al.\
(2009). However, we often found slight but noticeable offsets also in
FORS data for stars that are not linearly polarized. These offsets can
be naturally explained if the ratio between the transmission functions
in the perpendicular beam, and the transmission function in the
parallel beam, does not remain constant as the retarder waveplate is
set to the different position angles. The ratio between the
transmission functions can be estimated from the ratios of the \fo\
and \fe\ fluxes in each frame, and analytically propagated to the
final expressions of the \pv\ profiles using Eqs.~(3) of Bagnulo et
al. (2011).

Both \pv\ and \nv\ profiles show occasional spikes that occur in the
same wavelength bin. Most of these spikes are probably produced by
cosmic ray hits, and, if not removed, may lead to spurious detections
of magnetic fields, or at least decrease the precision of its
determination. As a remedy, we clip the \pv\ profiles by discarding
from the computation of the \bz\ values those points for which the
(rectified) \nv\ value departs from zero by more than $3\,\sigma$.

%%%%%%%%%%%%%%%%%%%%%%%%%%%%%%%%%%%%%%%%%%%%
\begin{figure}
\includegraphics[angle=270,width=16cm]{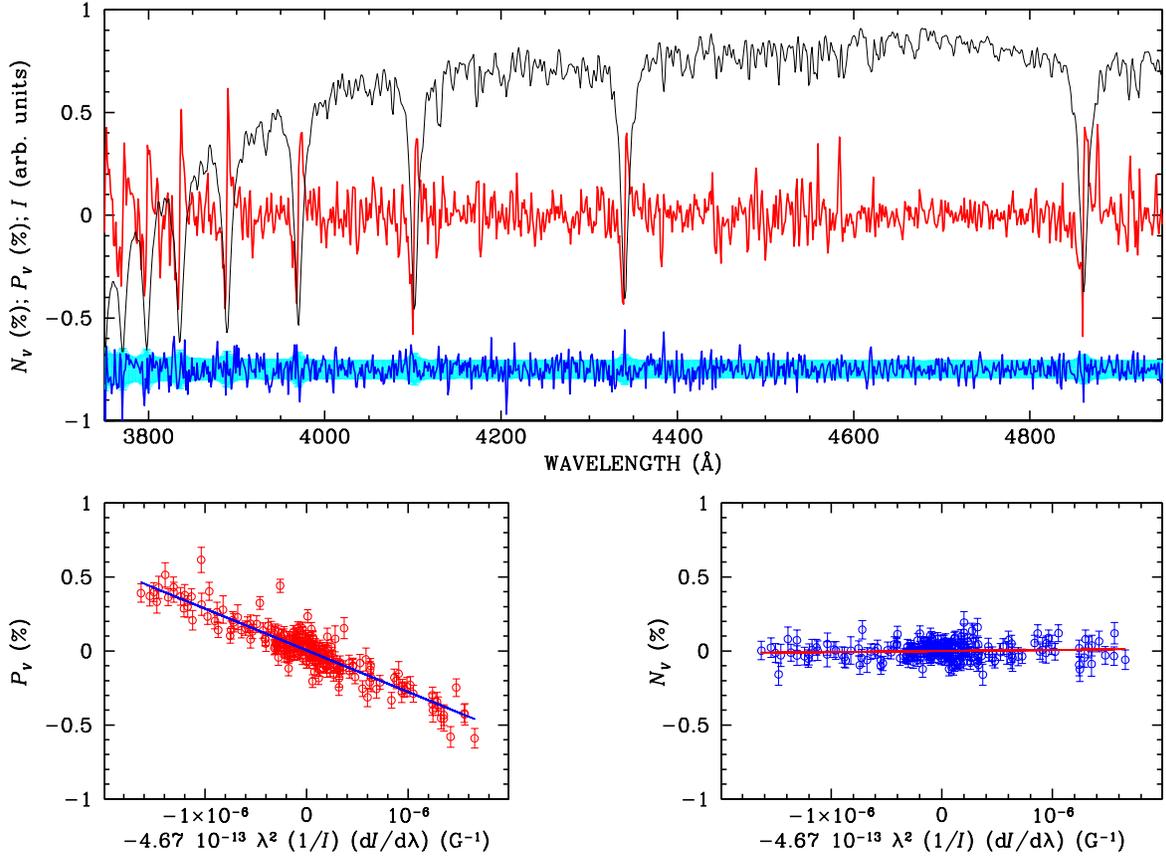}
\caption{\label{Fig_HD94660} 
The polarized spectrum of the magnetic Ap star HD\,94660.
The top panel shows Stokes $I$ (black solid line, in arbitrary units,
and not corrected for the instrument response), \pv\ (red solid line
centred about 0), and the null profile (blue solid line, offset by
-0.75\,\% for display purpose). Light blue bars represent the error bars
of \pv. The slope of the red interpolating lines in the bottom
panels gives the mean longitudinal field from \pv\ (left panel) and
from the null profile (right panel, the latter being expected
zero), both calculated using the H Balmer lines only. (After 
Bagnulo et al. 2011).
}
\end{figure}
%%%%%%%%%%%%%%%%%%%%%%%%%%%%%%%%%%%%%%%%%%%%
Figure~\ref{Fig_HD94660} shows the results obtained for the FORS
observations of the well known magnetic star HD\,94660.

\subsection{3.1 Errors bars}\label{Sect_Photon_Noise}
%%%%%%%%%%%%%%%%%%%%%%%%%%%%%%%%%%%%%%%%%%%%
\begin{figure}
  \includegraphics[angle=270,width=16cm]{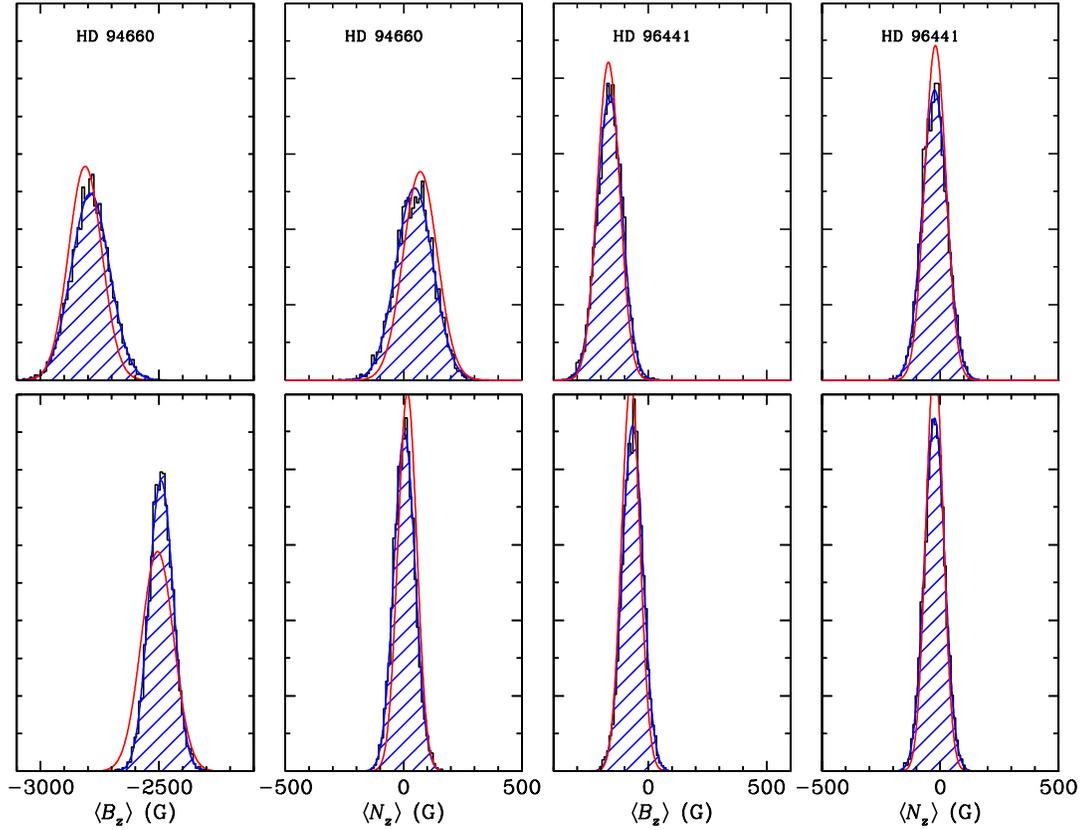}
  \caption{\label{Fig_MC} Monte Carlo simulations compared to predictions of the 
 ``standard'' error theory for a strongly magnetic star (HD\,94660) and a
non magnetic star (HD\,96441). Top panels refer to the \bz\ and \nz\ values obtained 
from the analysis of the H Balmer lines, and the bottom panels to the values obtained
from the analysis of the full spectrum. Blue shaded histograms show the distributions
of the \bz\ and \nz\ values obtained after Monte Carlo simulations. Red solid lines
show Gaussian curves centred on the original \bz\ and \nz\ values with $\sigma$
calculated via usual standard error propagation.}
\end{figure}
%%%%%%%%%%%%%%%%%%%%%%%%%%%%%%%%%%%%%%%%%%%%

The uncertainties of the \bz\ measurements can be calculated according
to a well established error theory; assuming that each individual step
of the measurement process can be linearised, one can propagate the
error bars of the measured flux to the \pv\ profile (see, e.g.,
Eqs.~(A3-A6) of Bagnulo et al.\ 2009), and then obtain the \sbz\ error
bar associated to the \bz\ measurement. However, since there are
indications that \bz\ error bars may be underestimated, we have
preferred to adopt the more conservative approach to multiply the
analytical expression for \sbz\ by the square root of the reduced
$\chi^2$.  The validity of this approach has been tested with Monte
Carlo simulations.  We have considered all raw frames associated to a
certain number of measurements, and calculated the \bz\ and \nz\
values and their error bars. Then we have re-considered the same
series of frames and scattered each pixel of each scientific frame
according to a Gaussian distribution with $\sigma$ equal to the error
pertaining to that pixel. We have re-reduced the so altered set of
frames, and measured the \bz\ and \nz\ values from the H Balmer lines
only and from the full spectrum. We have repeated this procedure 5000
times, and obtained for each set of frames two distributions of \bz\
values and two distributions of \nz\ values. We have compared these
distributions with the Gaussian curves centred on the \bz\ and \nz\
values measured from the unaltered frames, with $\sigma$ calculated
via standard propagation theory. The results for two specific cases are shown in
Fig.~\ref{Fig_MC}. The discrepancy observed for the case of \bz\
obtained from the analysis of the full spectrum of HD\,96441 (left
bottom panel of Fig.~\ref{Fig_MC}) may be due to the fact that our
error bar calculated analytically is weighted by a reduced
$\chi^2$. This value is definitely $>1$ because Eq.~(\ref{Eq_Bz}) is
not strictly valid for blended metal lines when the field is
strong. All the remaining cases shown in Fig.~\ref{Fig_MC} confirm
that \bz\ and \nz\ errors due to photon noise are correctly
calculated.

\subsection{3.2 Uncertainties not due to photon-noise}\label{Sect_Non_Photon_Noise}
%%%%%%%%%%%%%%%%%%%%%%%%%%%%%%%%%%%%%%%%%%%%
\begin{figure}
  \includegraphics[width=5cm]{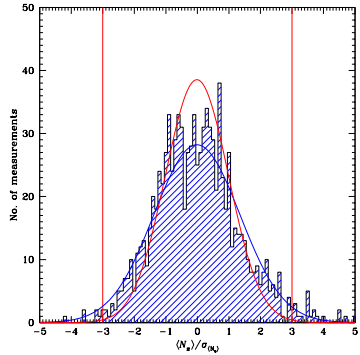}
  \caption{\label{Fig_Null_Fields} The actual distribution of the values \nz\ normalised
to their error bars (shaded histogram) compared to their theoretical distribution, i.e.,
a Gaussian with $\sigma=1$. Vetical bars show the $\pm 3\,\sigma$ limits. (After Bagnulo
et al.\ 2011).}
\end{figure}
%%%%%%%%%%%%%%%%%%%%%%%%%%%%%%%%%%%%%%%%%%%%

There are indications that error bars due to photon-noise
underestimate the \textit{actual} uncertainties.

As mentioned in Sect.~1, in the literature there are non negligible
discrepancies between the \bz\ values obtained by different authors
who analyse the \textit{same} FORS dataset. Note also that sometimes, 
the same authors have revised their own field determinations. This
is probably due to the fact that \bz\ measurements
depend, sometimes crucially, on the numerous specific choices taken
during data reduction.  For instance, the \bz\ and \nz\ values
depend on the algorithm adopted for spectral extraction, on how
data are clipped to remove the effects of cosmic rays, on whether
flat-field correction is applied or not. A complete description of all
these issues would be outside of the scope of anything but a detailed
monograph, hence here we limit ourselves to refer the reader to a few
numerical examples given in Bagnulo et al. (2011), and to warn that
changes in the data reduction algorithm may lead to change of the \bz\
values comparable to $1\,\sigma$.

A global inconsistency emerges from the analysis of the full FORS
archive dataset, even when all data are reduced by consistently
adopting the same algorithm, i.e., an unusually large number of field
detections in the null profiles. Figure~\ref{Fig_Null_Fields} shows
the distribution of the null field values normalised to their error
bars compared to the expected Gaussian with $\sigma=1$. This anomaly
could be explained by a number of reasons, e.g., (i) that for some
dataset, data reduction needs an extremely refined tuning; (ii) stars
change radial velocity during an observing series; (iii) during the
observations there are instrument flexures. It is likely that all
these issues play a role, but we should note that in particular,
instrument flexures, if present, would be completely out of user
control. A flexure as small as a quarter of a pixel cannot be detected
with the monitoring of the position of sky lines, because most of the
observations were obtained with very short exposure times, and because
in general the observed spectral range includes only one telluric
line. Yet, simple numerical simulations show that even such small
flexures may change the \bz\ and \nz\ values by an amount comparable
to that of the error bars due to photon-noise.  Occasional flexures of
the order of 1 pixel could be responsible for the observed high
frequency of \nz\ detections (about 3\,\%-5\,\%).

\section{4. Results}\label{Sect_Results}
The magnetic field measurements obtained with FORS1 have been directed at
three main goals: detection of previously unknown fields, measurements
of fields in stars already known to be magnetic, and establishing
upper limits to fields in stars which are not magnetic.  

Our discussion of reduction procedures for FORS1 spectropolarimetric
data shows that using different reduction procedures to infer magnetic
field strengths leads to different results. These include changes in
both the value of the mean longitudinal field \bz\ and the associated
uncertainty \sbz\ obtained from any specific observation.

This finding is probably not a serious concern for measurements of
fields at the tens or hundreds of $\sigma$ level, as these
measurements are much more sensitive to such factors as the lines
available in the spectrum, the spectral region used, and the degree of
chemical and inhomogeneity in the atmosphere.

On the other hand, underestimating the actual measurement uncertainty 
can have quite important effects when the issue is whether
or not a field has been detected at all in a particular star or class
of stars. Failure to recognize that computed uncertainties may not
fully represent the range of measurement values that may be taken by
\bz\ even in the absence of a real stellar field can lead to spurious
discoveries. These in turn can lead to much effort being expended to
analyse or explain the presence of fields that are not in fact there,
or at least have not been reliably detected. It is in this context
that the most harmful effects of incorrect measurements, or underestimated
measurement uncertainties, can occur.

Because of problems that spurious discovery announcements can cause in
a research field, we have carried out a systematic analysis of new
magnetic field detections obtained at less than about the $6 \sigma$
level. In the following we report on some of our findings. Our
complete revision of FORS1 magnetic data, which will include more
classes of stars, will be published in two forthcoming papers (Bagnulo
et al. 2011, and in preparation).

\paragraph{Herbig AeBe stars}
The first measurement of a magnetic field in a Herbig Ae star obtained
with the FORS1 instrument was reported by Hubrig et al. (2004), who
measured $\bz=-450\pm93$\,G in HD~139614. This discovery was
deemed as spurious detection by Wade et al.\ (2005) (who from the same
FORS1 data measured $\bz=-150\pm50$\,G), but re-affirmed as real by
Hubrig et al.\ (2006c). A new reduction by Hubrig et al.\ (2007a) led
to the revised value of $\bz=-112\pm36$\,G, and our re-reduction of the
discovery observation ($\bz=-84\pm65$\,G) confirms that no field is
detected in this measurement. The same star was observed several times
using the high-resolution spectropolarimeter ESPaDOnS (Wade et al.\
2005, 2007), and no magnetic field was ever detected.

The results of a FORS1 survey of 49 Herbig AeBe stars were described
by Wade et al. (2005; 2007).  Magnetic fields were reported in two
stars: HD~72106A (whose companion HD~72106B is a Herbig star), and
HD~101412. The uncertainties of the measurement of these two stars
have increased in our re-reduction of these data, but both fields are
still detected at $~ 3\,\sigma$ level. These discoveries are
corroborated by independent observations obtained with ESPaDOnS.
Wade et al. (2005; 2007) report also marginal detections in other three
Herbig stars, but these detections were never confirmed by subsequent
observations, nor by our new data reduction.

Further surveys of Herbig AeBe stars using FORS1 were described by
Hubrig et al. (2006c; 2007a), who reported the discovery of magnetic
fields in two new Herbig stars. Hubrig et al.\ (2009b) report new
magnetic field detections in six more Herbig stars (almost 30\,\% of
the observed sample). In our re-reduction of all these data, only
three measurements remain signficant. Detections in HD~144432 and in
HD~144668 (at 3\,$\sigma$ level) might be real, but require
confirmation. A detection at the $5 - 6 \sigma$ level of a field in
HD~150193 appears stronger in our re-reduction than originally
published. However, a single ESPaDOnS observation with an uncertainty
of about 15~G detects no field (E. Alecian, private communication). In
conclusion, most of FORS1 field discoveries of Herbig AeBe stars
appear to be spurious, and detectable magnetic fields appear to occur
in only $\le 10$\% of all Herbig AeBe stars observed with FORS1.
 
\paragraph{HgMn stars}  
Some of the peculiar A stars of the HgMn type, and in particular
$\alpha$~And = HD~358, have been repeatedly observed with FORS1.
Based on these data, Hubrig et al.\ (2006b) have claimed field
detections in four HgMn stars. 

Numerous investigations of magnetism in HgMn (e.g., Borra \& Landstreet
1980, Shorlin et al. 2002, Makaganiuk et al. 2011) have found no
convincing evidence for fields in this class of stars with
uncertainties often in the range of a few G up to a few tens of G.
Furthermore, $\alpha$~And has been specifically investigated with the
MuSiCoS (13 observations, typical $\sigma \sim 30 - 60$\,G) and
ESPaDOnS (5 observations, typical $\sigma
\sim 6 - 19$~G) spectropolarimeters without any significant detection
Wade et al.\ (2006).  Our new reductions have increased the standard errors
of all FORS1 measurements by factors of ~1.3, and none of our
results shows a field at the $3\,\sigma$ level.

\paragraph{$\beta$ Cephei pulsators and Slowly Pulsating B stars}
Hubrig et al. (2006a) and Hubrig et al.\ (2009a) have reported the
discovery of a field in the $\beta$~Cep star $\xi^1$~CMa = HD~46328,
based on altogether 13 detections at approximately the 6 or $7 \sigma$
level.  The field always appears to be close to $+350$~G. This
detection is supported by a single field detection by Silvester et
al.\ (2009) using ESPaDOnS at about the $30 \sigma$ level. Our
re-reduction of these data also confirms the presence of a
field. This discovery appears very robust, and clearly indicates the
capability of FORS1 to detect rather modest magnetic fields.

Major FORS1 surveys of both $\beta$~Cep and SPB stars have been carried
out by Hubrig et al. (2006a) and Hubrig et al.\ (2009a). Nearly 70 stars were
observed for magnetic fields using the FORS1 spectropolarimeter (19
known and suspected $\beta$~Cep stars, 50 known and suspected SPBs).
Fields were reported to have been detected in five
$\beta$~Cep stars and 26 SPB stars. If these results were confirmed
they would be quite important. Such high incidence of detected fields
would strongly suggest that magnetism is intrinsically connected with
the pulsation phenomenon in early B stars, as it is in the cool
rapidly oscillating Ap (roAp) stars.

However, all reported field detections in $\beta$~Cep stars, except
those of $\xi^1$~CMa, have mostly become insignificant in the new
reductions. Only two field measurements, one each for HD~74575 and
HD~136504, are still barely significant at the $3 \sigma$ level.  Of
the more than 40 field detections reported in SPB stars by Hubrig et
al. (2006a) and Hubrig et al.\ (2009a), all but five have decreased to
non-detections in the new reductions. Our reduction suggest that
fields might be present in HD~53921, HD~152511 (with two still
apparently significant detections), and HD~208057. However, some or
all of these detections could be manifestations of the occasional
outliers found in the FORS1 data. Note that a field has also been
marginally detected by Silvester et al. (2009) in a single ESPaDOnS
measurement of HD~208057.

Based on the re-reductions of the FORS1 data as well as other
published material, we conclude that as in other kinds of upper main
sequence stars, magnetic fields are relatively rare in $\beta$ Cep and
SPB stars. Furthermore, there is no significant case for considering
that the pulsation properties of these B pulsators are intrinsically
connected to the presence of weak magnetic fields.

\paragraph{O-type stars}
Hubrig et al.\ (2008) carried out the first large survey of O-type stars
with FORS1, and reported field measurements of 13 stars (including one
star previously reported by Hubrig et al.\ 2007b). Field discoveries
were claimed for five of these stars. Our reduction confirms a
marginal detection in only one of these stars, the O6.5f?p star
HD~148937. The field of this star has been detected in three more
observations with FORS2 by Hubrig et al.\ (2011) and with ESPaDONs
observations obtained by Wade et al.\ (2011). This discovery is clearly
real, although the field of this star, always about $-250$~G, is close
to the limit for reliable field detection by FORS1 for such stars.

\paragraph{Central stars of planetary nebulae}
Jordan et al.\ (2005) carried out six field measurements of four
central stars of planetary nebulae (CSPN), and concluded that at least
two of these stars have highly significant fields in the kG
range. This result supported the idea that such fields have an
important shaping effect on the the planetary nebulae themselves, and
suggest that there must have been important loss of magnetic flux
between this evolution state and the white dwarf state where stars
with as much flux as that inferred for the CSPN are relatively
rare. However, this result was called into question recently by Leone
et al.\ (2011), who obtained one new measurement of each of the CSPNs
NGC~1360 and LSS~1362, detecting no significant field in either star.
Leone et al.\ (2011) also re-reduced the older data of Jordan et al.\
(2005), finding no significant fields. Our new reductions are
completely consistent with the re-reductions carried out by Leone et
al.\ (2011).  None of the original field measurements of Jordan et
al.\ (2005) reveals a significant field. As remarked by Leone et al.\
(2011), there is now no significant evidence for coherent magnetic
fields in the central stars of planetary nebulae, and the current best
upper limits on \bz\ are at roughly the 1 to 2~kG level.

\section{5. Conclusions}\label{Sect_Conclusions}
FORS is an instrument perfectly capable of performing field
measurements of the order of at least 250--300\,G even in faint and
rapidly rotating stars. However, for weaker fields, it is well
possible that limits intrinsic to the technique, or even to the
instrument itself, are reached. In particular, we have shown that (i)
different data reduction techniques lead to
results that differ for values comparable to (and in some cases larger
than) photon-noise error bars; (ii) instrument flexures as small as a
quarter of a pixel may affect the field measurements by an amount
again comparable to a typical photon-noise error bar. Since FORS is
mounted at the Cassegrain focus, such small flexures might occasionally
occur. We conclude that, before being accepted as real, many
recent claims of weak field detections obtained with FORS in various
kinds of stars need to be confirmed by new and repeated measurements
obtained both with FORS itself and, independently, with other
instruments.

\begin{theacknowledgments}
Carlo Izzo was responsible for the pipelines of various instruments of
the La Silla-Paranal Observatory, including the FORS instrument.
Without his help and friendship this work would not have been
possible. Carlo died after short illness a few days before this
workshop. He will be deeply missed by many friends inside and outside
ESO.
\end{theacknowledgments}

\end{document}